\documentstyle[12pt,twoside]{article}

\def\a{\alpha}
\def\b{\beta}

\def\d{\delta}
\def\f{\phi}                    
\def\g{\gamma}

\def\j{\psi}
\def\k{\kappa}
\def\l{\lambda}
\def\m{\mu}
\def\n{\nu}

\def\p{\pi}                     
\def\r{\rho}                    

\def\F{\Phi}
\def\G{\Gamma}

\def\L{\Lambda}

\def\S{\Sigma}

\def\cf{{\cal F}}

\def\ch{{\cal H}}

\def\cl{{\cal L}}

\def\cs{{\cal S}}

\catcode`@=11

\def\un#1{\relax\ifmmode\@@underline#1\else $\@@underline{\hbox{#1}}$\relax\fi}

{\catcode`\'=\active \gdef'{{}~\bgroup\prim@s}}

\def\magstep#1{\ifcase#1 \@m\or 1200\or 1440\or 1728\or 2074\or 2488\or
        2986\fi\relax}

\font\twfvmi=cmmi10\@magscale5
    \skewchar\twfvmi='177
\font\twfvsy=cmsy10\@magscale5
    \skewchar\twfvsy='60
\font\twfvly=lasy10\@magscale5
\font\thtyrm=cmr10\@magscale6

\def\vpt{\textfont\z@\fivrm
  \scriptfont\z@\fivrm \scriptscriptfont\z@\fivrm
\textfont\@ne\fivmi \scriptfont\@ne\fivmi \scriptscriptfont\@ne\fivmi
\textfont\tw@\fivsy \scriptfont\tw@\fivsy \scriptscriptfont\tw@\fivsy
\textfont\thr@@\tenex \scriptfont\thr@@\tenex \scriptscriptfont\thr@@\tenex
\def\prm{\fam\z@\fivrm}%
\def\unboldmath{\everymath{}\everydisplay{}\@nomath
  \unboldmath\fam\@ne\@boldfalse}\@boldfalse
\def\boldmath{\@subfont\boldmath\unboldmath}%
\def\pit{\@getfont\pit\itfam\@vpt{cmti5}}%
\def\psl{\@subfont\sl\it}%
\def\pbf{\@getfont\pbf\bffam\@vpt{cmbx5}}%
\def\ptt{\@subfont\tt\rm}%
\def\psf{\@subfont\sf\rm}%
\def\psc{\@subfont\sc\rm}%
\def\ly{\fam\lyfam\fivly}\textfont\lyfam\fivly
    \scriptfont\lyfam\fivly \scriptscriptfont\lyfam\fivly
\@setstrut\rm}

\def\@vpt{}

\def\vipt{\textfont\z@\sixrm
  \scriptfont\z@\sixrm \scriptscriptfont\z@\sixrm
\textfont\@ne\sixmi \scriptfont\@ne\sixmi \scriptscriptfont\@ne\sixmi
\textfont\tw@\sixsy \scriptfont\tw@\sixsy \scriptscriptfont\tw@\sixsy
\textfont\thr@@\tenex \scriptfont\thr@@\tenex \scriptscriptfont\thr@@\tenex
\def\prm{\fam\z@\sixrm}%
\def\unboldmath{\everymath{}\everydisplay{}\@nomath
  \unboldmath\@boldfalse}\@boldfalse
\def\boldmath{\@subfont\boldmath\unboldmath}%
\def\pit{\@subfont\it\rm}%
\def\psl{\@subfont\sl\rm}%
\def\pbf{\@getfont\pbf\bffam\@vipt{cmbx6}}%
\def\ptt{\@subfont\tt\rm}%
\def\psf{\@subfont\sf\rm}%
\def\psc{\@subfont\sc\rm}%
\def\ly{\fam\lyfam\sixly}\textfont\lyfam\sixly
    \scriptfont\lyfam\sixly \scriptscriptfont\lyfam\sixly
\@setstrut\rm}

\def\@vipt{}

\def\xxxpt{\textfont\z@\thtyrm
  \scriptfont\z@\twfvrm \scriptscriptfont\z@\twtyrm
\textfont\@ne\twfvmi \scriptfont\@ne\twfvmi \scriptscriptfont\@ne\twtymi
\textfont\tw@\twfvsy \scriptfont\tw@\twfvsy \scriptscriptfont\tw@\twtysy
\textfont\thr@@\tenex \scriptfont\thr@@\tenex \scriptscriptfont\thr@@\tenex
\def\unboldmath{\everymath{}\everydisplay{}\@nomath\unboldmath
        \textfont\@ne\twfvmi \textfont\tw@\twfvsy \textfont\lyfam\twfvly
        \@boldfalse}\@boldfalse
\def\boldmath{\@subfont\boldmath\unboldmath}%
\def\prm{\fam\z@\thtyrm}%
\def\pit{\@subfont\it\rm}%
\def\psl{\@subfont\sl\rm}%
\def\pbf{\@getfont\pbf\bffam\@xxxpt{cmbx10\@magscale6}}%
\def\ptt{\@subfont\tt\rm}%
\def\psf{\@subfont\sf\rm}%
\def\psc{\@subfont\sc\rm}%
\def\ly{\fam\lyfam\twfvly}\textfont\lyfam\twfvly
   \scriptfont\lyfam\twfvly \scriptscriptfont\lyfam\twtyly
\@setstrut \rm}

\def\@xxxpt{}

\def\Huge{\@setsize\Huge{36pt}\xxxpt\@xxxpt}

\font\thtymi=cmmi10\@magscale6
    \skewchar\thtymi='177
\font\thtysy=cmsy10\@magscale6
    \skewchar\thtysy='60
\font\thtyly=lasy10\@magscale6
\font\thsirm=cmr12\@magscale6

\def\xxxvipt{\textfont\z@\thsirm
  \scriptfont\z@\thtyrm \scriptscriptfont\z@\twfvrm
\textfont\@ne\thtymi \scriptfont\@ne\thtymi \scriptscriptfont\@ne\twfvmi
\textfont\tw@\thtysy \scriptfont\tw@\thtysy \scriptscriptfont\tw@\twfvsy
\textfont\thr@@\tenex \scriptfont\thr@@\tenex \scriptscriptfont\thr@@\tenex
\def\unboldmath{\everymath{}\everydisplay{}\@nomath\unboldmath
        \textfont\@ne\thtymi \textfont\tw@\thtysy \textfont\lyfam\thtyly
        \@boldfalse}\@boldfalse
\def\boldmath{\@subfont\boldmath\unboldmath}%
\def\prm{\fam\z@\thsirm}%
\def\pit{\@subfont\it\rm}%
\def\psl{\@subfont\sl\rm}%
\def\pbf{\@getfont\pbf\bffam\@xxxpt{cmss12\@magscale6}}%
\def\ptt{\@subfont\tt\rm}%
\def\psf{\@subfont\sf\rm}%
\def\psc{\@subfont\sc\rm}%
\def\ly{\fam\lyfam\thtyly}\textfont\lyfam\thtyly
   \scriptfont\lyfam\thtyly \scriptscriptfont\lyfam\twfvly
\@setstrut \rm}

\def\@xxxvipt{}

\def\HUGE{\@setsize\HUGE{43pt}\xxxvipt\@xxxvipt}

\catcode`@=12

\font\tenex=cmex10 scaled 1200

\def\Sc#1{\hbox{\sc #1}}        

\def\bo{{\raise.05ex\hbox{\large$\Box$}\:}}             
\def\cbo{{\,\raise-.15ex\Sc [\,}}                       
\def\pa{\partial}                                       
\def\su{\sum}                                           
\def\TH{{\raise.2ex\hbox{$\displaystyle \bigodot$}\mskip-4.7mu \llap H \;}}
\def\face{\hbox{\normalsize$\;\;\:{\raise.9ex\hbox{\oo n}\mskip-13mu \llap
        {${\buildrel{\hbox{\frtnrm ..}}\over\smile}$}}\:$}}     
\def\Face{{\raise.2ex\hbox{$\displaystyle \bigodot$}\mskip-2.2mu \llap {$\ddot
        \smile$}}}                                      
\def\Lhat{{\bf\rlap{\kern-.09em$\hat{\phantom L}$}L}}
\def\Lcheck{{\bf\rlap{\kern-.09em$\check{\phantom L}$}L}}

\def\sp#1{{}^{#1}}                              
\def\sb#1{{}_{#1}}                              
\def\sl#1{\rlap{\hbox{$\mskip 1 mu /$}}#1}      
\def\sket#1{\left| #1\right\rangle}             
\def\leftrightarrowfill{$\mathsurround=0pt \mathord\leftarrow \mkern-6mu
        \cleaders\hbox{$\mkern-2mu \mathord- \mkern-2mu$}\hfill
        \mkern-6mu \mathord\rightarrow$}
\def\dvec#1{\vbox{\ialign{##\crcr
        \leftrightarrowfill\crcr\noalign{\kern-1pt\nointerlineskip}
        $\hfil\displaystyle{#1}\hfil$\crcr}}}           
\def\dt#1{{\buildrel {\hbox{\LARGE .}} \over {#1}}}     
\def\ddt#1{{\buildrel {\hbox{\LARGE .\kern-2pt.}} \over {#1}}}

\def\frac#1#2{{\textstyle{#1\over\vphantom2\smash{\raise.20ex
        \hbox{$\scriptstyle{#2}$}}}}}                   
\def\ha{\frac12}                                        
\def\sfrac#1#2{{\vphantom1\smash{\lower.5ex\hbox{\small$#1$}}\over
        \vphantom1\smash{\raise.4ex\hbox{\small$#2$}}}} 
\def\bfrac#1#2{{\vphantom1\smash{\lower.5ex\hbox{$#1$}}\over
        \vphantom1\smash{\raise.3ex\hbox{$#2$}}}}       
\def\afrac#1#2{{\vphantom1\smash{\lower.5ex\hbox{$#1$}}\over#2}}    

\def\boxes#1{
        \newcount\num
        \num=1
        \newdimen\downsy
        \downsy=-1.64ex
        \mskip-7.8mu
        \bo
        \loop
        \ifnum\num<#1
        \llap{\raise\num\downsy\hbox{$\bo$}}
        \advance\num by1
        \repeat}
\def\boxup#1#2{\newcount\numup
        \numup=#1
        \advance\numup by-1
        \newdimen\upsy
        \upsy=.82ex
        \mskip7.8mu
        \raise\numup\upsy\hbox{$#2$}}

\newskip\humongous \humongous=0pt plus 1000pt minus 1000pt
\def\caja{\mathsurround=0pt}

\newif\ifdtup
\def\panorama{\global\dtuptrue \openup2\jot \caja
        \everycr{\noalign{\ifdtup \global\dtupfalse
        \vskip-\lineskiplimit \vskip\normallineskiplimit
        \else \penalty\interdisplaylinepenalty \fi}}}
\def\li#1{\panorama \tabskip=\humongous                         
        \halign to\displaywidth{\hfil$\displaystyle{##}$
        \tabskip=0pt&$\displaystyle{{}##}$\hfil
        \tabskip=\humongous&\llap{$##$}\tabskip=0pt
        \crcr#1\crcr}}

\def\CMP{Commun. Math. Phys.}
\def\NP{Nucl. Phys. B}
\def\PL{Phys. Lett. }

\def\PRD{Phys. Rev. D}
\def\ref#1{$\sp{#1]}$}

\parskip=\medskipamount                 
\def\baselinestretch{1.2}       
\thicklines                         
\def\title#1#2#3#4{
\begin{document}
        {\hbox to\hsize{#4 \hfill Imperial-TP/ #3}}\par
        \begin{center}\vskip.5in minus.1in {\Large\bf #1}\\[.5in minus.2in]{#2}
        \vskip1.4in minus1.2in {\bf ABSTRACT}\\[.1in]\end{center}
        \begin{quotation}\par}
\def\author#1#2{#1\\[.1in]{\it #2}\\[.1in]}
\def\AM{Aleksandar Mikovi\'c\,
\footnote{E-mail address: A.MIKOVIC@IC.AC.UK}\footnote{Based on the
talks presented at Trieste conference on Gauge
Theories, Applied Supersymmetry and Quantum Gravity, May 1993
and at Danube '93 Workshop, Belgrade, Yugoslavia, June
1993}
\\[.1in] {\it Blackett Laboratory, Imperial
College,\\ Prince Consort Road, London SW7 2BZ, U.K.}\\[.1in]}

\def\endtitle{\par\end{quotation}\vskip3.5in minus2.3in\newpage}


\def\endabstract{\par\end{quotation}
        \renewcommand{\baselinestretch}{1.2}\small\normalsize}


\def\xpar{\par}                                         
\def\letterhead{
        \centerline{\large\sf IMPERIAL COLLEGE}
        \centerline{\sf Blackett Laboratory}
        \vskip-.07in
        \centerline{\sf Prince Consort Road, SW7 2BZ}
        \rightline{\scriptsize\sf Dr. Aleksandar Mikovi\'c}
        \vskip-.07in
        \rightline{\scriptsize\sf Tel: 071-589-5111/6983}
        \vskip-.07in
        \rightline{\scriptsize\sf E-mail: A.MIKOVIC@IC.AC.UK}}
\def\sig#1{{\leftskip=3.75in\parindent=0in\goodbreak\bigskip{Sincerely yours,}
\nobreak\vskip .7in{#1}\par}}


\def\ree#1#2#3{
        \hfuzz=35pt\hsize=5.5in\textwidth=5.5in
        \begin{document}
        \ttraggedright
        \par
        \noindent Referee report on Manuscript \##1\\
        Title: #2\\
        Authors: #3}


\def\start#1{\pagestyle{myheadings}\begin{document}\thispagestyle{myheadings}
        \setcounter{page}{#1}}


\catcode`@=11

\def\ps@myheadings{\def\@oddhead{\hbox{}\footnotesize\bf\rightmark \hfil
        \thepage}\def\@oddfoot{}\def\@evenhead{\footnotesize\bf
        \thepage\hfil\leftmark\hbox{}}\def\@evenfoot{}
        \def\sectionmark##1{}\def\subsectionmark##1{}
        \topmargin=-.35in\headheight=.17in\headsep=.35in}
\def\ps@acidheadings{\def\@oddhead{\hbox{}\rightmark\hbox{}}
        \def\@oddfoot{\rm\hfil\thepage\hfil}
        \def\@evenhead{\hbox{}\leftmark\hbox{}}\let\@evenfoot\@oddfoot
        \def\sectionmark##1{}\def\subsectionmark##1{}
        \topmargin=-.35in\headheight=.17in\headsep=.35in}

\catcode`@=12

\def\sect#1{\bigskip\medskip\goodbreak\noindent{\large\bf{#1}}\par\nobreak
        \medskip\markright{#1}}
\def\chsc#1#2{\phantom m\vskip.5in\noindent{\LARGE\bf{#1}}\par\vskip.75in
        \noindent{\large\bf{#2}}\par\medskip\markboth{#1}{#2}}
\def\Chsc#1#2#3#4{\phantom m\vskip.5in\noindent\halign{\LARGE\bf##&
        \LARGE\bf##\hfil\cr{#1}&{#2}\cr\noalign{\vskip8pt}&{#3}\cr}\par\vskip
        .75in\noindent{\large\bf{#4}}\par\medskip\markboth{{#1}{#2}{#3}}{#4}}
\def\chap#1{\phantom m\vskip.5in\noindent{\LARGE\bf{#1}}\par\vskip.75in
        \markboth{#1}{#1}}
\def\refs{\bigskip\medskip\goodbreak\noindent{\large\bf{REFERENCES}}\par
        \nobreak\bigskip\markboth{REFERENCES}{REFERENCES}
        \frenchspacing \parskip=0pt \renewcommand{\baselinestretch}{1}\small}
\def\unrefs{\normalsize \nonfrenchspacing \parskip=medskipamount}
\def\Item{\par\hang\textindent}
\def\Itemitem{\par\indent \hangindent2\parindent \textindent}
\def\makelabel#1{\hfil #1}
\def\topic{\par\noindent \hangafter1 \hangindent20pt}
\def\Topic{\par\noindent \hangafter1 \hangindent60pt}

\title{NON-PERTURBATIVE TWO-DIMENSIONAL DILATON GRAVITY}
{\AM}{92-93/44}{July 1993}
We present a review of the canonical quantization approach to the
problem of non-perturbative 2d dilaton gravity. In the case of chiral matter we
describe a method for solving the constraints by constructing a
Kac-Moody current algebra. For the models of interest, the relevant Kac-Moody
algebras are based on $SL(2,{\bf R})\otimes U(1)$ group and on an
extended 2d Poincare group. As a consequence, the constraints become
free-field Virasoro generators with background charges.
We argue that the same happens in the non-chiral case. The problem of
the corresponding BRST cohomology is discussed as well as the
unitarity of the theory. One can show that the theory is unitary by
chosing a physical gauge, and hence the problem of transitions from
pure into mixed sates is absent. Implications for the physics of black
holes are discussed.
\endtitle

\sect{1. Introduction}

\noindent Two-dimensional dilaton gravity theories of interest can be described
by an action
$$ \li{ S &= S_0 + S_m \cr
        S_0 &=  \int_{M} d^2 x \sqrt{-g} e^{-2\F}\left[ R +
\g (\nabla \F)^2 + U(\F ) \right]\cr
S_m &= -\ha\int_M d^2 x\sqrt{-g} \su_{i=1}^N (\nabla\f_i)^2 &(1.1)\cr}$$
where $\F$ and $\f_i$ are scalar fields, $\g$ is a constant, $g$, $R$
and $\nabla$ are
determinant, scalar curvature and covariant derivative respectively,
associated with a metric on the 2d manifold $M$. For our purposes we
will assume that $M= \S \times {\bf R}$, so that $\S = S^1$ (a circle) or
$\S = {\bf R}$ (a real line). We will
refer to these two cases as compact and non-compact respectively.

$S_0$ describes the
coupling of the dilaton $\F$ to the metric, while $S_m$ represents
conformally coupled scalar matter. Depending on the value of the
constant $\g$ and the form of the potential $U$, one can get various
dilaton gravity theories. The relevant examples are:

(1) Spherical symmetry reduction of the 4d Einstein-Hilbert action \cite{napi}
$$ S_0 =  \int_{M} d^2 x \sqrt{-g} e^{-2\F}\left[ R +
2(\nabla \F)^2 + \k e^{2\F}  \right] \quad.\eqno(1.2)$$
The reduction is performed by decomposing the 4d line element as
$$ ds^2 = g_{\m\n}dx^\m dx^\n + e^{-2\F(t,r)}(d\theta^2 +
\sin^2\theta d\f^2 )\quad, \eqno(1.3)$$
where $(r,\theta,\f)$ are the spherical coordinates, $x^\m = (t,r)$, while
$\k$ is the inverse Newton constant.

(2) Dimensional reduction of 4d dilaton gravity \cite{rev}
$$ S_0 =  \int_{M} d^2 x \sqrt{-g} e^{-2\F}\left[ R +
4(\nabla \F)^2 + 4\l^2  \right]\quad. \eqno(1.4)$$

(3) Induced 2d gravity
$$ S_0 =  \int_{M} d^2 x \sqrt{-g}\left[
\ha(\nabla \f)^2 + \a R\f + \L  \right]\quad, \eqno(1.4)$$
which is not of the form (1.1) but it can be related to it by a field
redefinition \cite{tsey}.

{}From these examples it is clear that 2d dilaton gravity theories can
be relevant as toy models for 4d quantum gravity and for non-critical
string theory. Although model (1) contains a black hole solution by
definition, it was an interesting discovery that model (2)
contains a black hole solution as well \cite{wit}. Furthermore, when
(2) is coupled to scalar conformal matter, it is exactly solvable
\cite{cghs}. Given the fact that it is also a renormalizible field
theory, this makes it an excellent toy model for the study of the
black hole evaporation and backreaction. Semi-classical analysis of
the model in the limit of large $N$ had raised a hope that a black
hole singularity can dissapper in the quantum theory \cite{cghs}.
However, very soon it was shown that a singularity is present in the
solution of the semi-classical equations of motion \cite{2dbh}.
Furthermore, Hawking
has given general arguments for the existence of singularity in
any semi-classical approximation \cite{hawk2}. All this indicates that
one should perform a non-perturbative, more precisely exact,
quantization of the theory in order to see what really happens with
the black hole.

\sect{2. Non-perturbative approaches}

\noindent Non-perturbative formulation of any quantum theory of gravity has to
deal with the following conceptual problems, which do not appear in
ordinary field theories

(1) no background metric

(2) maintaining diffeomorphism invariance

(3) problem of time

(4) space-time singularities.

\noindent For a more detailed discussion see \cite{ish}.
The non-perturbative approaches to 2d dilaton gravity which have been
studied so far are path-integral and canonical. The idea of the path-integral
approach is to perform the functional integral over the metric,
dilaton and matter fields exactly, and then to study the corresponding
effective action and the correlation functions (for a review and
references see \cite{{rev},{gid}}). Beside its own
difficulties in achieveing the stated goal,
it is not clear how to construct the physical Hilbert
space whithin this approach and how to address the corresponding
conceptual questions.

In the canonical approach \cite{{mik1},{mik2},{uch},{hir},{ver}}
the construction of the physical Hilbert
space is the primary goal, from which all other questions are
answered. This is achieved from the study of the constraints, which
can be derived by using the ADM (Arnowit, Deser, Misner) method
\cite{{mik1},{bil}}. The ADM method
takes care of the problems (1) and (2), as opposed to the method used
in \cite{{hir},{ver}}, where the constraints were derived in the
conformal gauge, and the quantization was based on the space of
classical solutions.

Before explaining the canonical ADM formulation, we will briefly study
field redefinitions, in order to arrive at the simplest possible form of
the action. That in turn simplifies the constraints. Let $\j^2 =
e^{-2\F}$, then $S_0$ from the eq. (1.1) becomes
$$ S_0 =  \int_{M} d^2 x \sqrt{-g}\left[
\ha(\nabla \j)^2 +{1\over 2\g} R\j^2 + \tilde{U}(\j)\right]\quad,
\eqno(2.1)$$
where $\j$ has been rescaled into ${1\over\sqrt{2\g}}\j$ ($\g \ne 0$).
Then by performing a Russo-Tseytlin transformation \cite{tsey}
$$ \f = {1\over \g} \j^2 \quad,\quad \tilde{g}_{\m\n}=
e^{-2\r}g_{\m\n} \quad,\quad 2\r = {1\over \g} \j^2 - {\g\over 2}\ln\j
\eqno(2.2)$$
we get
$$ S_0 =  \int_{M} d^2 x \sqrt{-\tilde{g}}\left[
\ha(\tilde{\nabla} \f)^2 + \ha \tilde{R}\f +V(\f)\right]\quad, \eqno(2.3)$$
where $V(\f)= \tilde{U} e^{2\r}$. In the dilaton gravity case $V =
\ha\l^2 e^\f$, and hence consider
$$ S_0 =  \int_{M} d^2 x \sqrt{-g}\left[
\ha(\nabla \f)^2 + \a R\f + \L e^{\b\f}\right]\quad, \eqno(2.4)$$
where $\a,\b$ and $\L$ are constants.
The action (2.4) represents a class of solvable dilaton gravity
theories, which can be seen by redefining the metric as $\tilde{g}_{\m\n}
= e^{\b\f}g_{\m\n} $ \cite{mik1}, so that
$$ S_0 =  \int_{M} d^2 x \sqrt{-\tilde{g}}\left[
\ha(1-2\a\b)(\tilde{\nabla} \f)^2 + \ha \tilde{R}\f + \L \right]\quad.
\eqno(2.5)$$
Since eq. (2.5) represents an induced gravity action, one can use
Polyakov's results about the existence of an $SL(2,{\bf R})$ current
algebra \cite{poly} to solve the theory. In the canonical setting this
can be done by constructing gauge-independent currents, which form an
$SL(2,{\bf R})\otimes U(1)$ current algebra \cite{{mik1},{abd}}. However, in
\cite{mik1} it was not realised that for $\a\b = \ha$, i.e.
precisely in the dilaton gravity case, this current algebra degenerates into
an extended 2d Poincare current algebra \cite{uch}. As we are going to
show, the
construction proposed in \cite{mik2} for solving the theory works even in
that case, after slight modifications.

\sect{3. Canonical formulation}

\noindent Consider the following action
$$ S_0 = - \int_{M} d^2 x \sqrt{-g}\left[
{\g\over 2}(\nabla \f)^2 + \a R\f + V(\f)
+\ha\su_{i=1}^N (\nabla \f_i)^2 \right]\quad, \eqno(3.1)$$
where $\a$ and $\g$ are constants. Note that the field redefinitions
of the previous section always scaled the metric so that the form of
the matter action is unchanged, because of its conformal invariance.
Canonical reformulation requires $M = \S \times {\bf R}$, and it
simplifies if we use the ADM parametrization of the metric
$$g_{\m\n}=\pmatrix{-{\cal N}^2 + gn^2 & gn \cr gn & g \cr} \quad,\eqno(3.2)$$
where $\cal N$ and $n$ are the laps function and the shift vector respectively,
while $g$ is a metric on $\S$. By defining the canonical momenta as
$$ p = {\pa \cl\over \pa \dt{g}}\quad,\quad \p = {\pa \cl\over \pa
\dt{\f}} \quad,\quad \p_i = {\pa \cl\over \pa \dt{\f_i}} \quad, \eqno(3.3)$$
where $\cl$ is the Lagrange density of (3.1) and dots stand for $t$
derivatives, the action becomes
$$ S= \int dt dx \left( p\dt{g} + \p\dt{\f} + \p^i\dt{\f_i} -
{{\cal N}\over\sqrt{g}}G_0 - n G\sb 1 \right) \quad. \eqno(3.4)$$
The constraints $G_0$ and $G_1$ are given as
$$\li{G\sb 0 (x) =&  - {\g\over2\a^2}(gp)^2
+{1\over\a}gp\p + {\g\over2}(\f^{\prime})^2 + gV(\f)
- 2\a \sqrt{g}\left( {\f^{\prime}\over\sqrt{g}}\right)^{\prime}\cr & +
\ha \su_{i=1}^N (\p_i^2 + (\f_i^{\prime})^2) \cr
G\sb 1 (x) =& \p\f^{\prime} - 2p^{\prime}g - pg^{\prime}+
\su_{i=1}^N \p_i\f^{\prime}_i \quad,&(3.5)\cr}$$
where primes stand for $x$ derivatives.
The $G$'s generate the diffeomorphisms of $M$, such that $G_1$ generates the
diffeomorphisms of $\S$, while $G_0$ generates time-translations of $\S$.
A special feature of two dimensions is that
$$ T_{\pm} = \ha (G_0 \pm G_1) \eqno(3.6)$$
generate two commuting copies of the one-dimensional diffeomorphism
algebra. When $\S = S^1$ these become two commuting Virasoro algebras.

As in the 4d canonical gravity, direct quantization of the constraints
(3.5) is problematic due to their non-polynomial dependence on the
canonical variables. One way around this problem is to follow the
strategy introduced by Ashtekar in the 4d case \cite{asht}, which is
to find new canonical variables such that the constraints become polynomial.
This can be done by constructing first a Kac-Moody current algebra
corresponding to the hidden symmetries we discussed in
the previous section \cite{{mik1},{uch}}. Let us introduce the
following variables
$$\li{J\sp + &= -{1\over{g}} T\sb - + {\L\over 2}\cr
J\sp 0 &= gp + \d \left( \p + \a\,
{g^{\prime}\over g}\right) + (\a + \g\d)\f^{\prime} \cr
J\sp - &= 4\a^2 g \quad,\cr
P_D &=  {1\over\sqrt2}\left(\p + \a\, {g^{\prime}\over g} +
\g\f^{\prime} \right) \quad.&(3.7)\cr}$$
When $\g \ne 0$, then for $\d = -{\a\over\g}$ $(J^a , P_D)$ form an
$SL(2,{\bf R})\otimes U(1)$ Kac-Moody algebra
$$\li{ \{ J^0 (x), J^\pm (y)\} &= \pm J^\pm (x) \d (x-y) \cr
  \{ J^+ (x), J^- (y)\} &= -2\g J^0 (x) \d(x-y)+4\a^2\d^{\prime}(x-y)  \cr
\{ J^0 (x), J^0 (y)\} &= -{2\a^2\over\g}\d^{\prime}(x-y) \cr
\{ P_D (x) , P_D (y) \} &= \g \d^{\prime}(x-y)\quad,&(3.8)\cr}$$
where all other Poisson brackets are zero.
When $\g =0$ then for $\d=0$ one gets a centrally extended Poincare
Kac-Moody algebra
$$\li{ \{ J^0 (x), J^\pm (y)\} &= \pm J^\pm (x) \d (x-y) \cr
       \{ J^+ (x), J^- (y)\} &= 2\sqrt2\a P_D (x) \d (x-y) + 4\a^2
\d^{\prime}(x-y)  \cr
\{ J^0 (x), P_D (y)\} &= 2\sqrt2\a\d^{\prime}(x-y) \cr
\{ P_D (x) , P_D (y) \} &= 0 \quad,&(3.9)\cr}$$
and all other Poisson brackets are zero.
The importance of this centrally extended Poincare group for the
dilaton gravity was first recognized in \cite{jack}.

The next step is to perform a generalized Sugawara construction, which
gives a generator of 1d diffeomorphisms quadratic in the Kac-Moody currents
$$ T_G = {1\over4\a^2}[J^+J^- - \g (J^0)^2] - (J^0)^{\prime} +
{\a + \g\d\over\sqrt2\a^2}J^0 P_D - {\d(2\a +\g\d)\over2\a^2}P_D^2 + \d
\sqrt2 P_D^{\prime} \,.\eqno(3.10)$$
Then it is straightforward to show that
$$ T_G + T_M^+ = G_1 \quad,\eqno(3.11)$$
where
$$T_M^+ = \ha\su_{i=1}^{N} (P_i^+)^2\quad,\quad P_i^{\pm}
={1\over\sqrt2}(\p_i \pm \f_i^{\prime}) \quad.\eqno(3.12)$$
Eq. (3.11) implies that we can replace the constraints $G_0 =0$ and
$G_1=0$ with
$$J^+ - \ha\L = 0 \quad,\quad T_G + T_M^+ = 0 \quad.\eqno(3.13)$$

Given the new set of constraints (3.13), one could quantize the current
algebra $(J^a, P_D, P_i^+, P_i^-)$ and then try to find the physical Hilbert
space by using the representation theory of the Kac-Moody algebras.
However, this method works only in the case of chiral matter ($P_i^- =
0$) since
$$\{ J^+ (x), P_i^- (y)\} = -{1\over g (x)}P_i^-(x)\d^{\prime}(x-y)
\quad.\eqno(3.14) $$
This is not a big restriction, since
the chiral matter case is relevant for a one-sided collapse.

Instead of using the group-theoretical method, we look for a further
change of variables
such that we get a truly canonical variables. This approach
is more
convenient for addressing the problem of time and the
unitarity of the theory \cite{mik2}. In the $SL(2)$ case there exists
a change of
variables, called Wakimoto construction \cite{wak}, which can be
written classically as
$$ \li{J^+ &= B\cr
J^0  &=-B\G + \a\sqrt2 P\sb L \cr
J^-  &= B\G^2 -2\a\sqrt2 \G P\sb L  - 4\a^2 \G^{\prime} \quad,&(3.15)\cr}$$
where the new fields satisfy
$$ \{B (x) ,\G (y) \} =  \d (x-y) \quad,\quad
\{ P\sb L (x) , P\sb L (y)\} = - \d^{\prime} (x-y)
\quad,\eqno(3.16)$$
and all other Poisson brackets are zero. In the Poincare case the
analog of the eq. (3.16) is
$$ \li{J^+ &= B\cr
J^0  &=-B\G + 2\a Y^{\prime}  \cr
J^-  &= 2\sqrt2\a \G P_D - 4\a^2 \G^{\prime}
\quad,&(3.17)\cr}$$
where the non-zero Poisson brackets are
$$ \{ B (x) ,\G (y) \} =  \d (x-y) \quad,\quad
\{ P_D(x) , Y (y)\} = -\d (x-y)
\quad.\eqno(3.18)$$
The generalized Sugawara tensor (3.10) takes the following form
$$T_G = B^{\prime}\G - \ha P^2_0  + \ha P^2_1 +
Q_0 P^{\prime}_0  + Q_1 P^{\prime}_1\eqno(3.19)$$
where $P_0 = P_L$ and $P_1 = P_D$ in the $SL(2)$ case, while
$P_0 = {1\over\sqrt2}(P_D - Y^{\prime})$ and $P_1 = {1\over\sqrt2}
(P_D + Y^{\prime})$ in the Poincare case.
The background charges $Q_0$
and $Q_1$ are given as $(-\sqrt2\a,-\sqrt2\a)$ in the $SL(2)$ case or
$(\sqrt2\a,-\sqrt2\a)$ in the Poincare case. Note that by performing a
canonical transformation
$$ P_0 = -\tilde{P}_0 \quad,\quad P_1 = \tilde{P}_1 \eqno(3.20)$$
one can transform the Poincare Sugawara tensor into the $SL(2)$ one.

By defining a new canonical pair $(P,X)$, where
$$ P = B - \ha\L  \quad,\quad
   X = \G \quad,\eqno(3.21)$$
the constraints now read
$$ P = 0 \eqno(3.22)$$
and
$$ \cs =  - \ha P^2_0  + \ha P^2_1 +
Q_0 P^{\prime}_0 +  Q_1 P^{\prime}_1
+ \ha \su_{i=1}^N (P_i^+)^2  = 0 \quad.\eqno(3.23)$$
By fixing the spatial diffeomorphism invariance as
$$ X(x) = x \eqno(3.24)$$
$P$ is eliminated by eq. (3.22), and therefore we are left with
only one constraint (eq. (3.23)) for the variables
$(P_0,P_1)$ and $( \p_i , \f_i )$. The $\cs$ constraint will play the
role of the Wheeler-DeWitt equation.

In the non-chiral case ($P_i^- \ne 0$), the constraints $T_+$ and
$T_-$ can be recasted in the form (3.23), which follows from the study
of the space of the classical solutions \cite{{hir},{ver}}. However,
the explicit canonical transformation which achieves that has not
been yet constructed in the ADM formalism.

\sect{4. Quantization}

\noindent In the canonical approach, there are two basic ways of
quantizing a constrained system

(1) quantize first and then solve the constraints (Dirac quantization),

(2) solve the constraints first and then quantize (reduced phase space
    (RPS) $\quad\quad$ quantization).

\noindent The Dirac quantization, and its variations (Gupta-Bleuler
and BRST method), are often preferred to the RPS quantization
because of preservation of the manifest symmetries of the theory. On the
other hand, RPS quantization is easier to accomplish. In our case, we
have followed so far the RPS method, and we will continue to do so,
but given the special role the constraints play in gravity, we will also
explore the Dirac quantization.

In order to accomplish the Dirac quantization of the $\cs$ constraint,
we need $\S$ to be compact, since otherwise we do not know much about the
representations of the 1d diffeomorphism algebra.
This creates a problem for the non-compact case, i.e. the one
where the black hole solutions exist, and this is usually resolved
by putting the system into a large box, of length $L$.

Now we will label the vector $(P_0, P_1, P_i^+)$ as $P_I$, so that
$$P_I (x) = {1\over\sqrt{L}}\left( p_I + \su_{n\ne 0} \a_n^I e^{in\p
x/L}  \right)\eqno(4.1)$$
where $p_i = 0$. Then
$$\li{ S(x) = & {1\over L}\su_n L_n e^{in\p x/L} \cr
        L_n =& \ha\su_m \a_{n-m}^I \a_m^I + in Q_I \a_n^I \quad,&(4.2)\cr}$$
where $Q_i = 0$. The $L_n$'s are promoted into operators acting on a
Fock space $\cf(\a^i_n)$ made out of the $\a_n$ modes in the standard way.
The $L_n$'s form a Virasoro algebra classically, but
in the quantum case there is an anomaly in the algebra, in the form of
the central extension term with the central charge $c = N$. This type
of situation is best handled in the BRST formalism. One
enlarges the original Fock space $\cf (\a^i_n)$ by introducing a canonical
pair of ghost fields $(b,c)$, and constructs a nilpotent operator
$$ \hat{Q} = \su_n c_{-n}( L_n - a\d_{n,0}) +
\ha\su_{n,m}(n-m):c_{-n}c_{-m}b_{n+m} : \quad.\eqno(4.3)$$
The nilpotency of $\hat{Q}$ requires
$$ -Q_0^2 + Q_1^2 = 2-N/12 \quad,\quad a = N/24 \quad,\eqno(4.4)$$
which is satisfied for $N=24$. The physical Hilbert space $\ch^*$ is
determined as the cohomology of $\hat{Q}$
$$\ch^* = Ker\,\hat{Q}/Im\,\hat{Q} \quad.\eqno(4.5)$$

The cohomology problem of this type has been studied extensively in
the $N=0$ case \cite{liz}. The physical Hilbert space has
three sectors
$$\li{\ch^* =& \ch_0^* \oplus \ch_1^* \oplus \ch_{-1}^* \cr
      \ch_0^* =& \{ \sket{p_0,p_1}\, |\, -p_0^2 + p_1^2 =0 \}\oplus\{ {\rm
discrete\,\, states}\} \cr
        \ch_{\pm 1}^* =& \{ {\rm discrete\,\, states} \} \quad,&(4.6)\cr}$$
where the subscripts $0,1,-1$ refer to the ghost number.
The $N\ne 0$ cohomology problem has not been studied in detail, but
one can deduce the following features. There will be only a zero-ghost
sector, since the intercept $a\ne 0$. A basis for the physical states
will be
$$      \ch_0^* = \{\, \sket{p_0,p_1}\otimes\a_{-n_1}^{i_1}...
\a_{-n_k}^{i_k}\sket{0} \, |\, -p_0^2 + p_1^2 + 2\su_{i=1}^k n_i =
N/12,\,\,  n_i>0\,\}\quad, \eqno(4.7)$$
i.e. the transverse oscilator states. The exact cohomology analysis
may say something about the discrete states, but we do not expect any
other continuous momentum states but the ones from the eq. (4.7).

This is confirmed by the results of the RPS quantization \cite{mik2}. Given the
$\cs$ constraint (eq. (3.23)), and the fact that $P_0$ and $P_1$ can be
always represented as
$$ P_0 = {1\over\sqrt2}(P_T - T^{\prime})\quad,\quad
P_1 = 2\sqrt2\a p + {1\over\sqrt2}(P_T + T^{\prime})\quad,\eqno(4.8)$$
where we have introduced an extra zero mode $p$ ($p^{\prime}=0$) and
$(P_T,T)$ is a canonical pair, one gets
$$ (2\a p + P_T)(2\a p + T^{\prime}) -2\a P_T^{\prime} +
\ha\su_{i=1}^N (P_i^+)^2 = 0 \quad.\eqno(4.9)$$
In the Poincare case one has to perform the canonical transformation
(3.20) first, in order to get the eq. (4.9). Now we chose $T$ as the
time variable, and at the same time we fix the diffeomorphism invariance
by a gauge choice
$$ T(x,t) = t \quad.\eqno(4.10)$$
We can now solve eq. (4.9) for $P_T$
$$ P_T (x) = -2\a p + {1\over4\a}\,e^{px}\left( k + \int^x dy e^{-py}
\su_{i=1}^N (P_i^+)^2 \right)\quad,\eqno(4.11) $$
where $k$ is the constant of integration.
Hence the independent canonical variables are $(p,q)$ and
$(\p_i,\f_i)$, which proves our conjecture in the Dirac quantization
that only the transverse mode states are physical\footnote{This is
true if there is no anomaly, i.e. $N=24$}.

We now proceed by specifying the Hamiltonian associated
with the gauge choice (4.10), which can be determined as
$$H = -\int_{\S} dx P_T(x) = c_1 p + {1\over4\a p}\left(\su_n \a_{-n}^i\a_n^i +
c_0 \right)\quad. \eqno(4.12)$$
The constants $c_0$ and $c_1$ can be determined from the boundary conditions,
although $c_0$ can have a quantum contribution due to the normal
ordering effects.

One gets an anologous formula to eq. (4.12) in the non-chiral case,
which follows from the fact that $T_+$ and $T_-$ can be recasted in
the form (3.23). Then
$$G_0 = T_+ + T_- = \ha \p_I^2 + \ha (\f_I^{\prime})^2 + \sqrt2 Q_I
\p_I^{\prime}\eqno(4.13) $$
and
$$G_1 = T_+ - T_- = \p_I \f_I^{\prime} + \sqrt2 Q_I
\f_I^{\prime\prime}\quad.\eqno(4.14)$$
By chosing the bosonic string light-cone gauge
$$ \p_+ = p \quad,\quad \f_+ = t \quad,\eqno(4.15)$$
where $V_{\pm} = V_0 \pm V_1$, one gets
$$H = {1\over2 p}\left(\su_n ( \a_{-n}^i\a_n^i + \tilde{\a}_{-n}^i
\tilde{\a}_n^i ) +
c_0 \right)\quad. \eqno(4.16)$$

The unitarity of the theory follows from the fact that the expression
(4.16) can be promoted into a Hermitian operator acting on the
physical Hilbert space
$$\ch^* = L^2(p)\otimes \cf (\a^i_n) \otimes \cf (\tilde{\a}_n^i )
\quad,\eqno(4.17) $$
where $L^2(p)$ is the Hilbert space of square-integrable functions of
$p$, while $\tilde{\a}$ are the modes of $P_i^-$.
Therefore one has a unitary evolution described by a Schrodinger
equation
$$ i{\pa\over\pa t} \Psi [t,p,\f_i(x)] = \hat{H}\Psi
[t,p,\f_i (x)] \quad,\eqno(4.18) $$
$\Psi \in \ch^*$, and hence no transitions from pure into mixed states
occur in this theory.

\sect{5. Conclusions}

\noindent The most straightforward conclusion is that the black holes in
the quantum theory
defined by eq. (4.18) do not destroy information, and
a unitary S-matrix can be constructed. The authors of \cite{ver} have
arrived at the
similar conclusions by studying canonical quantization of the dilaton
gravity in the conformal gauge. After imposing a reflecting boundary
condition, they construct a
S-matrix and prove its unitarity, without using the
Hamiltonian, which is a more difficult but an alternative way of proving the
unitarity.

The evolution of the matter is governed by a free-field Hamiltonian,
which is not surprising given the fact that the exact classical
solution can be written in terms of free fields \cite{cghs}.
Actually, this classical solvability of the theory is the reason why
it is possible to find a canonical transformation which makes the
constraints quadratic. The
non-trivial part of the Hamiltonian comes from its dependence on the zero-mode
$p$, which is the remnant of the graviton and dilaton degrees of freedom.

Although the quantum theory is unitary,
it is less clear what happens to the black hole.
This would require studying the operators associated with the
metric and the scalar curvature. The difficulty is that
$g$ and $R$ are complicated (non-polynomial) functions of the
free-field variables, and hence it is a non-trivial task to promote
them into well-defined Hermitian operators. Provided that this problem
is resolved, one could study the evaporation of the black holes in the
following way \cite{mik2}. Let $\Psi_0$ be a physical state at $t=0$ such that
$$ <\Psi_0|\hat{g}(x)|\Psi_0> = f(x)\quad,\quad
<\Psi_0|\hat{R}(x)|\Psi_0> = h(x) \eqno(5.1)$$
are regular functions for every $x\in \S$. Then
$\Psi(t) = e^{-i\hat{H}t}\Psi_0$
and for $t>t_0$ a horizon will form in the effective metric
$<\Psi(t)|\hat{g}(x)|\Psi(t)>$. A density matrix $\hat{\r}$ could
be calculated by tracing out the states which are beyond the horizon.
Then one could try to find out under what conditions $\hat{\r}$ takes
approximately the thermal form
$$ \hat{\r} \approx {1\over Z}e^{-\b \hat{H}} \eqno(5.2)$$
and what are the corrections to the Hawking temperature
$$ \b = {4\p\over\l} + ... \quad.\eqno(5.3)$$
One can also formulate the problem of computing the temperature
corrections in the S-matrix formalism \cite{ver}.

In spite of all these advances in formulating the exact quantum theory,
we think that the conceptual problem of the
space-time singularity is still unresolved. Although we managed to
find observables\footnote{An observable in this context is a quantity
which has weakly vanishing Poisson brackets with the constraints}
which are well defined at the singularity,
there will be
other observables, those associated with the scalar curvature, which will
not be well defined at the singularity. One way of resolving this problem
\cite{mik2} is to study
$$ R_{eff}(x,t) =
<\Psi_0|e^{i\hat{H}t}\hat{R}(x)e^{-i\hat{H}t}|\Psi_0>\quad. \eqno(5.4) $$
If it stays a regular function for every $x$ and $t$ and for every
$\Psi_0$  that satisfies
the conditions of eq. (5.1), then we could say that the singularity
has been removed from the quantum theory. A very similar idea has been
proposed in \cite{smo}. However, there is no
a priori reason for something like this to happen, and this
issue has to be a subject of further studies.

\end{document}